\documentclass[aps,prb,twocolumn,longbibliography,superscriptaddress,article]{revtex4-2}
\usepackage{epsfig}
\usepackage{epstopdf}
\usepackage{amsmath}
\usepackage{amsfonts}
\usepackage{amssymb}
\usepackage{hyperref}
\usepackage{bm}
\usepackage{makecell}
\usepackage{rotating}
\usepackage{hyperref}
\usepackage{multirow}
\usepackage{graphicx}
\usepackage{float}

\usepackage{graphicx}
\usepackage{dcolumn}
\usepackage{bm}
\usepackage{color}

\usepackage{tikz,xcolor,hyperref}

\definecolor{lime}{HTML}{A6CE39}
\DeclareRobustCommand{\orcidicon}{%
	\begin{tikzpicture}
	\draw[lime, fill=lime] (0,0)
	circle [radius=0.16]
	node[white] {{\fontfamily{qag}\selectfont \tiny ID}};
	\draw[white, fill=white] (-0.0625,0.095)
	circle [radius=0.007];
	\end{tikzpicture}
	\hspace{-2mm}
}

\foreach \x in {A, ..., Z}{%
	\expandafter\xdef\csname orcid\x\endcsname{\noexpand\href{https://orcid.org/\csname orcidauthor\x\endcsname}{\noexpand\orcidicon}}
}

\begin{document}

\title{Effect of polar distortions on the linear and nonlinear \\ anomalous Hall conductivity of altermagnetic $\alpha$-MnTe}

\author{Mathews Benny\orcidM}
\email{mbenny@magtop.ifpan.edu.pl}
\affiliation{International Research Centre Magtop, Institute of Physics, Polish Academy of Sciences, Aleja Lotnik\'ow 32/46, PL-02668 Warsaw, Poland}

\author{Sahar Izadi Vishkayi\orcidS}
\affiliation{School of Quantum Physics and Matter, Institute for Research in Fundamental Sciences (IPM), P. O. Box 19395-5531, Tehran, Iran}

\author{Amar Fakhredine\orcidF}
\affiliation{Institute of Physics, Polish Academy of Sciences, Aleja Lotnik\'ow 32/46, 02668 Warsaw, Poland}

\author{Chanchal K. Barman}
\email{barman@ifpan.edu.pl}
\affiliation{Institute of Physics, Polish Academy of Sciences, Aleja Lotnik\'ow 32/46, 02668 Warsaw, Poland}

\author{Carmine Autieri\orcidB}
\email{autieri@magtop.ifpan.edu.pl}
\affiliation{International Research Centre Magtop, Institute of Physics, Polish Academy of Sciences,
Aleja Lotnik\'ow 32/46, PL-02668 Warsaw, Poland}

\date{\today}
\begin{abstract}
Altermagnetic $\alpha$-MnTe with N\'eel vector along the $y$-axis exhibits a finite anomalous Hall conductivity (AHC) and weak ferromagnetism along the $z$-axis. As already demonstrated in the bulk, there is a breaking of the C$_6$ symmetry by the in-plane N\'eel vector, leaving a C$_2$-type magnetic symmetry. The (001) surface of $\alpha$-MnTe breaks the C$_2$, leaving only a time-reversed mirror symmetry with respect to the $x=0$ plane. Therefore, we demonstrate that, on the surface, the interplay between the breaking of crystal symmetry and N\'eel vector orientation reduces the space group from hexagonal P6$_3$/mmc to orthorhombic Amm2. As a result, the surface exhibits not only a polar distortion along the $z$-axis but also along the $y$-axis. To describe the surface of MnTe in an accessible way, we simplify the problem and examine the effect of the in-plane electric field in bulk MnTe. Moreover, as a doped ionic semiconductor, the properties of MnTe can be influenced by lattice polarization under an applied electric field. We investigate the interplay between the intrinsic anomalous Hall effect and lattice polarization, showing that polarization effects can substantially affect the AHC. Since the electric field breaks inversion symmetry, this contribution from the lattice polarization coexists with the nonlinear anomalous Hall effect, highlighting the rich transport phenomenology of altermagnets.
\end{abstract}


\maketitle
    
\section{Introduction}

In altermagnets, sites with opposite spin are connected by rotational symmetries, either proper or improper, and symmorphic or nonsymmorphic, but are not connected by translation or inversion symmetries\cite{Smejkal22beyond,doi:10.1126/sciadv.aaz8809,hayami2019momentum,hayami2020bottom,Smejkal22,Mazin2021,yuan2023degeneracy,Samanta2025,Sun2025,Xu2025,wei2024crystal,Zhang2025,D3NR03681B,D3NR04798A,ssxp-gz9l,Cuono23orbital,D4NR04053H,https://doi.org/10.1002/adfm.202505145,Berritta2025,jr65-4273,ref1,10.1088/1361-6633/ae8868}. While the breaking of time-reversal symmetry and weak ferromagnetism induced by spin-orbit coupling (SOC) have been known for several decades\cite{DZYALOSHINSKY1958241}, one of the most striking features of altermagnets is non-relativistic spin–momentum locking with even-wave symmetry. Non-relativistic spin–momentum locking refers to a phenomenon in which the electron spin orientation becomes locked to its crystal momentum due to rotational crystal symmetries. This locking ensures a symmetry-protected zero net magnetization in the non-relativistic limit. 

In systems with Kramers degeneracy, SOC preserves time-reversal symmetry and cannot generate a net magnetization. In altermagnetic compounds with broken time-reversal symmetry, however, SOC can generate both weak ferromagnetism and anomalous Hall effect\cite{PhysRevLett.130.036702,PhysRevB.111.184407}. The presence of altermagnetic spin splitting is therefore a necessary condition for the emergence of both weak ferromagnetism and the anomalous Hall effect. Depending on the crystal point group, different classes of altermagnets can exhibit anomalous Hall effect depending on the orientation of the Néel vector, defined as the difference between the spin vectors at the two magnetic sites and specifying the spin direction\cite{839n-rckn}. The simplest form of antisymmetric exchange is the staggered Dzyaloshinskii–Moriya interaction\cite{DZYALOSHINSKY1958241}, which produces relativistic weak ferromagnetism in altermagnets always orthogonal to the Néel vector\cite{PhysRevB.111.054442}. When SOC is included, the non-relativistic spin–momentum locking is inherited by the dominant spin component, while the other two components are referred to as subdominant. Beyond weak ferromagnetism, SOC can modify the spin–momentum locking, e.g., by inducing $d$-wave spin–momentum locking in the subdominant components. To describe this generalization, we introduce the term relativistic spin–momentum locking (RSML) to denote spin–momentum locking across all three spin components\cite{Fakhredine26,PhysRevB.109.024404,PhysRevB.110.144412}.
In the case of breaking of inversion symmetry, altermagnets can exhibit the Rashba effect, which can transform the altermagnet from $d$-wave to $p$-wave if the Rashba has one nodal plane in common with the spin-mometum locking of the altermagnet\cite{D6MH00357E,gong2026symmetryprotectednodalplanesaccidental,v3fg-6smc}

The nonlinear anomalous Hall effect can emerge in materials possessing a finite Berry curvature dipole (BCD). A BCD originates from an asymmetric distribution of Berry curvature in momentum space, typically enabled by inversion-symmetry breaking. Consequently, the magnitude of the nonlinear anomalous Hall response is directly determined by the strength of the BCD \cite{https://doi.org/10.1002/qute.202100056}. Due to rotational symmetry constraints, the altermagnets exhibit nontrivial quantum geometry even in the case of vanishing net-magnetizations\cite{PhysRevB.111.174407}.
Since the imaginary component of the quantum geometric tensor corresponds to the Berry curvature, the time-reversal symmetry-broken electronic band structure in altermagnets can induce substantial Berry curvature dipoles when inversion symmetry is absent \cite{mukherjee2025electricfieldcontrolledsecondorder}.
In contrast, centrosymmetric altermagnets do not support a second-order anomalous Hall response arising from BCD. Nevertheless, a third-order anomalous Hall effect can emerge through higher-order moments of Berry curvature and quantum metric \cite{PhysRevLett.133.106701,xiang2026thirdorderintrinsicanomaloushall}. Nonlinear effects have also been observed in the anomalous Edelstein response in altermagnets\cite{sxyb-5rcy}.

Altermagnetic MnTe has attracted considerable interest in recent years due to its rich electronic, magnetic, \cite{PhysRevLett.130.036702,Amin2024,k36v-91br,uykur2026revisitingsymmetryopticalphonons,yklc-9n6t,bb6r-2nwz,chen2026strainengineeringintrinsicanomalous,sarkar2026anomaloushalleffectsiliconcompatible} and topological properties.
In the NiAs-type crystal structure, MnTe is classified as a non-relativistic g-wave altermagnet, which transitions to a $d$-wave altermagnet in the relativistic limit\cite{Fakhredine26,hirakida2025multipoleanalysisspincurrents}, allowing noncollinear spin current\cite{Song2026}. Experimentally realized NiAs-type crystal structures of MnTe (or $\alpha$-MnTe) are intrinsically $p$-doped with the Fermi level at the top of the valence band. MnTe exhibits AHC\cite{PhysRevLett.130.036702} accompanied by a finite net magnetization\cite{kluczyk2023coexistence}, which is predominantly of orbital origin\cite{g32j-hnvz}. 
Furthermore, effects of Sb- and I-doping in $\alpha$-MnTe have been proposed as potential routes for tuning its electronic and magnetic properties\cite{k36v-91br}.
MnTe is also known to exhibit several polymorphisms, including wurtzite, which lacks spatial inversion symmetry. More recently, signatures of spatial inversion symmetry breaking have been reported even in the bulk samples of putative centrosymmetric MnTe\cite{wu2025opticalsignaturesnoncentrosymmetricstructural,ren2026atomicscaleobservationsymmetrybreaking}. In addition, Rashba-type spin splitting has been reported at the surface due to an electric field along the $z$-axis\cite{zeng2025nonaltermagneticspintexturemnte}.
A recent study\cite{ren2026atomicscaleobservationsymmetrybreaking} further revealed that the dominant Mn-site distortions condense into the lower-symmetry phases Cmc2$_1$ and Amm2, both of which exhibit polar displacements of Mn-sites along the $y$-axis.


While most ferromagnets exhibiting AHC are good metals, the family of altermagnets displaying AHC also includes doped ionic and polar semiconductors \cite{https://doi.org/10.1002/advs.202307306}. In ferromagnetic metals, the lattice polarization is typically absent, whereas it is an intrinsic property of ionic semiconductors. Consequently, there has been growing interest in recent years in studying AHC in systems where lattice polarization plays an essential role.
In ionic semiconductors, lattice polarization arises from the
relative displacement of positive and negative ions within the crystal
unit cell. Such displacements may occur in response to an external electric
field, the excitation of polar optical phonons, or the presence of
intrinsic spontaneous or piezoelectric polarization, as found in
non-centrosymmetric crystal structures.
The macroscopic polarization $\mathbf{P}$ arising from the lattice response can be expressed as the vector:
\begin{equation}
\mathbf{P} = \frac{1}{V} \sum_i Z_i^{*} \, \mathbf{u}_i ,
\end{equation}
where $V$ is the unit-cell volume, $Z_i^{*}$ denotes the Born effective
charge of ion $i$, and $\mathbf{u}_i$ is its displacement from the
equilibrium position. Lattice polarization is therefore intrinsically linked to atomic motion. The cations and anions shift in opposite directions,
creating an electric dipole moment within the unit cell. In materials exhibiting spontaneous lattice polarization, these ionic displacements persist even in the absence of an external electric field, giving rise to multiferroicity when combined with magnetic order \cite{Wang2025}.
Although $\alpha$-MnTe is typically hole-doped and exhibits finite electrical conductivity, the carrier concentration can vary across the sample. As a result, different parts of the samples can be more insulating and can be susceptible to the effect of the electric field.

In this work, we study the consequence of breaking rotational symmetries at the (001) surface of MnTe and investigate the influence of electric field-induced lattice polarization in this ionic semiconductor.
While the electric field slightly affects the overall band structure, the linear AHC is found to be very sensitive and can be affected even by a small amount of lattice polarization.

The paper is organized as follows: Sec. II is devoted to the rise of the in-plane electric field on the surface of MnTe. In Sec. III, we present the results for the collinear magnetic phase both in the absence and in the presence of an applied electric field. The computational details are provided in the appendix. Finally, Sec. IV summarizes the main findings and presents the conclusions.

\section{Surface reconstruction in M\lowercase{n}T\lowercase{e}}

\begin{figure}
    \centering
    \includegraphics[width=1\linewidth,angle=0]{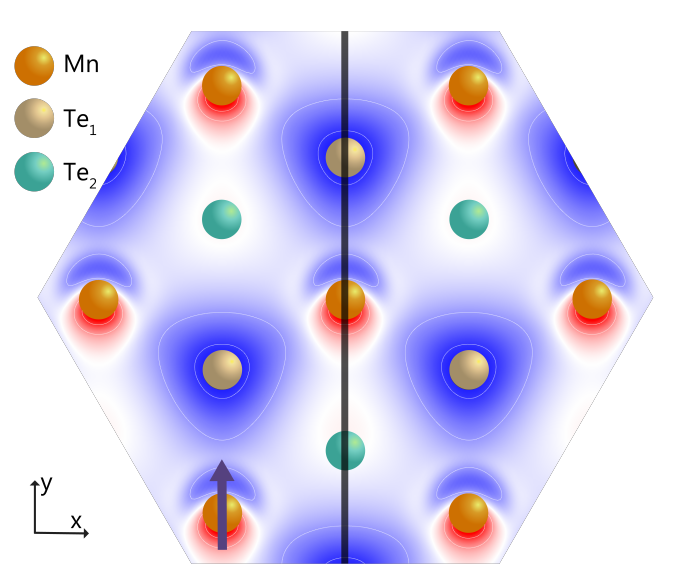}
    \caption{The charge density difference ($\Delta\rho = \rho_{\text{SOC}} - \rho_{\text{NOSOC}}$) evaluated within the $xy$-plane containing the surface Mn atoms. The prominent asymmetry in the accumulation (red) and depletion (blue) of electron density around the Mn centers demonstrates a clear breaking of the pristine threefold (C$_3$) lattice symmetry. This electronic reconstruction provides a direct visual signature of the SOC-driven in-plane structural polarization along the $y$-axis. The Te$_1$ and Te$_2$ atoms refer to tellurium atoms in two different layers above and below the Mn atoms. The purple arrow indicates the N\'eel vector orientation, and the solid vertical black line denotes the intersection of the vertical mirror plane, $M_x$, with the surface. The time-reversed vertical mirror plane, $\mathcal{T}M_x$, remains as a valid magnetic symmetry operator in this configuration. As a consequence, there is a reduction of the space group from P6$_3$/mmc to Amm2.}
    \label{fig:surface_reconst}
\end{figure}
In bulk $\alpha$-MnTe, which crystallizes in the P6$_3$/mmc space group (no. 194), each Mn atom is located at the center of an inversion-symmetric octahedron coordinated by six nearest-neighbour Te atoms. Experimentally, the most common surface to obtain is the one orthogonal to the z-axis, which we name the (001) surface.
To investigate the effect of (001) surface reconstruction, we consider a MnTe bilayer as a representative model; qualitatively similar behavior is also found for slabs containing an even number of layers.
Owing to the reduced symmetry and inequivalent magnetic environments at the surface, MnTe exhibits ferrimagnetism due to unequal spin moments at different lattice sites \cite{D3NR03681B}. This prediction has recently been corroborated by experimental research \cite{zhao2026emergentanomaloushalleffect,zhou2026surfacestatedrivenanomaloushalleffect}.

\begin{figure*}
    \centering
    \includegraphics[width=0.99\linewidth,angle=0]{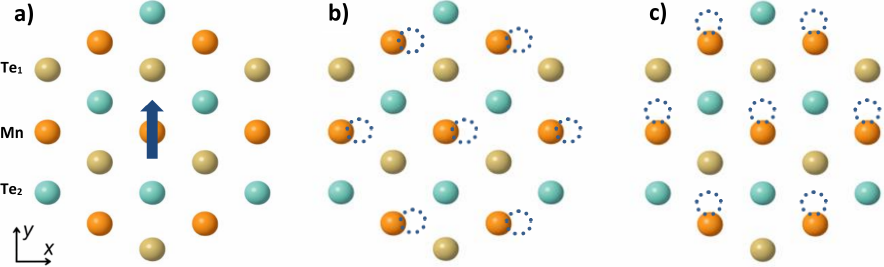}
    \caption{a) Top view of the MnTe crystal structure with centrosymmetric Space Group 194 (P6$_3$/mmc) without an electric field. The blue arrow indicates the N\'eel vector along the $y$-axis. b) Effect of lattice polarization induced by an electric field orthogonal to the N\'eel vector, resulting in a lattice polarization of $P_x \neq 0$. The dotted circles indicate the positions of the Mn atoms under the application of an electric field. The crystal symmetry is reduced to space group 40 (Ama2, orthorhombic). c) Effect of an electric field applied parallel to the N\'eel vector, resulting in a lattice polarization of $P_y \neq 0$. The crystal symmetry becomes space group 38 (Amm2, orthorhombic). There are three inequivalent atoms, which are Mn, Te$_1$, and Te$_2$; atoms Te$_1$ and Te$_2$ belong to different planes with different $z$-values.}
    \label{fig:Lattice_polarization}
\end{figure*}

In the unreconstructed MnTe bilayer, the crystal symmetry remains hexagonal, although the space group is reduced to P$\overline{6}$m2 (no. 187). We model the bilayer slab with symmetric Te terminations on both upper and lower surfaces, yielding an overall stoichiometry of Mn$_2$Te$_3$ within the unit cell arranged in a quintuple-layer Te-Mn-Te-Mn-Te stacking sequence. To eliminate artificial inter-slab interactions along the out-of-plane direction, a vacuum spacing of $27$ {\AA} is introduced.

Upon structural relaxation in the absence of SOC, the $z$-coordinates of the relaxed structure are modified, resulting in a reduction of the bilayer thickness. This contraction along the $z$-coordinates arises from surface and vacuum effects that lower the total energy of the system.
Consequently, the resulting bilayer after non-relativistic relaxation remains hexagonal belonging to the P$\overline{6}$m2 space group, which exhibits a threefold rotation axis (C$_3$) parallel to the $z$-axis and preserves a horizontal mirror plane. The preservation of these symmetries ensures that the Mn atoms remain centered within their local coordination environment and hence no in-plane displacements of the Mn sublattices are observed. We note that in this system the magnetic Mn atoms with opposite spins are related by a mirror symmetry rather than by a roto-translation. Nevertheless, the mirror symmetry is sufficient to enforce zero net magnetization. Furthermore, in the transition from a three-dimensional altermagnet to its two-dimensional counterpart, the Kramers' degeneracy cannot be restored in this system. Therefore, we have zero magnetization protected by mirror symmetry and non-relativistic spin-splitting in the altermagnetic fashion.

Upon inclusion of SOC in the presence of the N\'eel vector oriented along the $y$-direction, the C$_3$ rotational symmetry breaks down, which is consistent with the behavior earlier reported in bulk MnTe\cite{g32j-hnvz}.
As a consequence, the magnetic point group symmetry is reduced, and the lifting of C$_3$ symmetry forces the lattice to undergo an orthorhombic distortion during structural relaxation, where the magnetic space group transitions from P$\overline{6}$m2 to Amm2 as shown in Table \ref{tab:space_groups}. This distortion can also be characterised in terms of the magnetic layer group of the bilayer transitioning from P3m to Pm'.
This distortion displays a relative in-plane displacement of approximately 0.003 {\AA} along the $y$-direction between Mn and Te sublattices. For this magnetic configuration, a time-reversed mirror plane remains as the only symmetry operation. The relative displacement of Mn and Te sublattices generates a localized dipole moment, producing a local electric field at the surface. 
We highlight that both Mn and Te atoms have all coordinates in high-symmetry positions in $\alpha$-MnTe; therefore, no displacement is allowed without breaking the bulk symmetry. This shift of 0.003 {\AA} is truly an indicator of the breaking of inversion symmetry and it does not appear without spin-orbit coupling when the C$_3$ symmetry is preserved.



To visualize the impact of relativistic effects, we computed the electronic charge density difference of the MnTe bilayer, defined as $ \Delta \rho =\rho_{SOC} - \rho_{NOSOC} $, in the $xy$-plane containing the surface Mn atoms, as displayed in Fig. ~\ref{fig:surface_reconst}. In the absence of SOC, the crystal field environment around the Mn atoms preserves the threefold rotational symmetry (C$_3$). If this symmetry were maintained under SOC, the charge density difference would display a perfectly symmetric threefold rotational symmetry around each Mn atom. Instead, our results reveal asymmetric charge accumulation (red) and depletion (blue) profiles along the $y$-axis on opposite sides of Mn atoms. This anisotropic redistribution of charge density provides direct evidence of a SOC-induced electronic reconstruction influenced by structural atomic displacements.
These observations demonstrate that SOC, when coupled with Mn magnetic moments oriented along the $y$-direction, lowers the crystal symmetry. The resulting reduction in symmetry allows a lattice distortion that displaces the Mn atoms along the $y$-direction in the relaxed structure.

In the following discussion, we assume that the N\'eel vector remains oriented along the $y$-axis at the surface. However, this is not guaranteed since the energy difference between the N\'eel vector along $x$ and $y$ is quite small. Consequently, the surface magnetic order could in principle favor the N\'eel vector along the $x$-direction. Also, for the $x$-axis, there is an electric field along the $y$-axis. The effect of breaking inversion symmetry along the $y$-axis is a consequence of breaking the C$_3$ symmetry and an electric field along the $z$-axis. This same mechanism may also be relevant in the recently discovered polar phase of MnTe, which also exhibits polar distortions along the $y$-axis\cite{ren2026atomicscaleobservationsymmetrybreaking}. Indeed, these results yield magnetic space group 38.190 when the N\'eel vector is in the ab plane  (Amm'2', orthorhombic); this is consistent with the data reported in recent experiments \cite{ren2026atomicscaleobservationsymmetrybreaking}.

\begin{table}[h!]
\centering
\begin{tabular}{|c|c|c|}
\hline
\textbf{MnTe}  & \textbf{SG} & \textbf{MSG} \\
\hline
Bilayer ($\mathbf{N}\parallel y$) & $P\bar{6}m2$ (187) & $Amm'2'$ (38.190) \\
\hline
Bilayer + SOC ($\mathbf{N}\parallel y$)& $Amm2$ (38) & $Amm'2'$ (38.190) \\
\hline
Bilayer + SOC ($\mathbf{N}\parallel x$)& $Amm2$ (38) & $Amm'2'$ (38.190) \\
\hline
Bulk ($\mathbf{N}\parallel y$) & $P6_3/mmc$ (194) & $Cm'c'm$ (63.462) \\
\hline
Bulk + $E_x$ ($\mathbf{N}\parallel y$)& $Ama2$ (40) & $Ama2.1$ (40.203) \\
\hline
Bulk + $E_y$ ($\mathbf{N}\parallel y$)& $Amm2$ (38) & $Amm'2'$ (38.190) \\
\hline
\end{tabular}
\caption{Crystallographic space groups (SG) and magnetic space groups (MSG) of the MnTe systems considered in this work. $\mathbf{N}$ is the N\'eel vector defined as the difference between the spins on the two Mn atoms in the unit cell.}
\label{tab:space_groups}
\end{table}

The symmetry analysis of the bilayer and the bulk system with polar distortions along the $y$-direction reveal that they both share the same magnetic space group, Amm'2' as shown in Table \ref{tab:space_groups} along with the same magnetic layer group Pm'. This justifies the use of the bilayer as a representative system to study the surface effects of bulk MnTe. It also establishes the significance of the AHC calculations performed on bulk MnTe taking polar distortions into consideration.
In the following Section, we focus exclusively on the effect of the electric field on the anomalous Hall effect in bulk MnTe. A comprehensive investigation of the relativistic spin-momentum locking and anomalous Hall effect in freestanding MnTe will be presented in a separate publication\cite{BennyGong27}. 


\section{Anomalous Hall effect}


In Sec. \ref{subsec:hamil}, we describe the construction of the electronic Hamiltonian, starting from the non-magnetic phase and proceeding to the collinear altermagnetic phase of MnTe, both in the absence and in the presence of an applied electric field. In Sec. \ref{subsec:AHC}, we investigate the AHC of the collinear altermagentic state of MnTe. Finally, we examine the influence and impact of lattice polarization on the AHC.



\subsection{Construction of the Hamiltonian with and without electric field}
\label{subsec:hamil}


We begin by performing density functional theory calculations for the non-magnetic phase of $\alpha$-MnTe. Unlike the altermagnetic phase, the non-magnetic phase exhibits metallic behavior. In the corresponding electronic band structure, the Mn $d$-bands and Te $p$-bands predominantly lie within the energy range between -5.0 eV and +2.5 eV with respect to the Fermi level ($E_F$=0.0 eV), and they are well disentangled from the other bands. 
This makes the wannierization procedure much simpler with respect to the relativistic altermagnetic case reported in the literature\cite{g32j-hnvz}.

\begin{figure}
    \centering
    \includegraphics[width=0.99\linewidth]{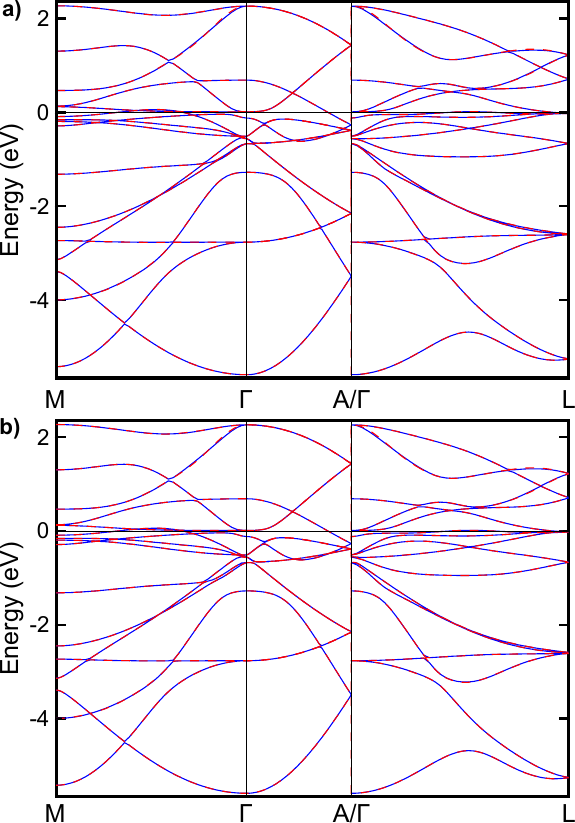}
    \caption{Non-magnetic band structures calculated by density functional theory on a distorted crystal structure with atomic displacements corresponding to an in-plane electric field, and their Wannierization.
    The DFT band structure is shown as a solid blue line, while the Wannier Hamiltonian is plotted with a dashed red line. (a) Band structure with electric field orthogonal to the N\'eel vector. (b) Band structure with electric field along the N\'eel vector. The zero of the energy scale is set at the top of the valence band.}
    \label{fig:DFT_NM}
\end{figure}

 To investigate the influence of lattice polarization on the AHC of $\alpha$-MnTe, we consider the response of the crystal to an applied electric field. In transport measurements, the conductivity relates the electric current density to the resulting electric field. While the internal electric field is efficiently screened in good metals, finite electric fields can persist in the doped semiconductors, such as MnTe. Beyond linear response, an electric field can induce ionic displacements, causing cations and anions to shift in opposite directions and thereby generating lattice polarization. 
 To simulate the consequences of an applied electric field in $\alpha$-MnTe, we introduce a rigid displacement of the Mn atoms along the $x$- and $y$-directions, corresponding to an electric field applied along the $x$- and $y$-axes, respectively. Such an approach was termed as the so-called “lattice-mediated” method, in which field-induced polar ionic displacements are used to mimic the structural response of the crystal to an externally applied electric field\cite{bousquet2023sign}. 
 The resulting lattice-polarized structures are illustrated in Fig.~\ref{fig:Lattice_polarization}. To construct the Wannier Hamiltonian, Wannierization is then performed for the non-magnetic phase in two distinct distorted structures: one with Mn shifted along the $x$-axis [Fig.~\ref{fig:Lattice_polarization}(b)] and the other with shifts along the $y$-axis [Fig.~\ref{fig:Lattice_polarization}(c)].
The in-plane Mn--Mn distance is 4.108 {\AA}, while the imposed Mn atom coordinate displacement is 0.0123 {\AA} (0.3\% of the lattice constant). This order of magnitude is comparable to the atomic displacements typically found in ferroelectric materials.
The non-magnetic Hamiltonians for the electric field along the $x$- and $y$-axis are named H(E$_x$) and H(E$_y$), respectively. The electronic band structures obtained from Hamiltonian H(E$_x$) and H(E$_y$) are reported in Fig. \ref{fig:DFT_NM}. 
The wannierization yields an almost perfect agreement between the DFT band structure and the corresponding Wannier Hamiltonian. From a visual inspection of the band structures [Fig.\ref{fig:DFT_NM}(a) and (b)], the differences between the two configurations are barely discernible.
However, the symmetry reduction depends on the direction of the electric field and is more pronounced for $E_x$ than for $E_y$. This is because the crystallographic distortions are perpendicular to the N\'eel vector for $E_x$, whereas these distortions are parallel to the N\'eel vector for E$_y$. Consequently, only the M$_z$ time-reversed mirror plane survives for distortions along $x$, while for distortions along $y$, both M$_x$ and M$_z$ time-reversed mirror planes remain intact. 
The qualitative difference between the electric field along the $x$-axis and the electric field along the $y$-axis is the presence of additional hopping parameters as t$^{001/2}_{xy,yz}$, t$^{001/2}_{xy,x^2-y^2}$ and t$^{001/2}_{xy,3z^2-r^2}$, arising when the electric field is along the $x$-axis, where the hopping parameters t$^{lmn}_{\alpha,\beta}$ are defined as between the orbitals $\alpha$ and $\beta$ between Mn sites connected by the vector l$\bold{a}$+m$\bold{b}$+n$\bold{c}$ where $\bold{a}$, $\bold{b}$ and $\bold{c}$ are the lattice vectors and l,m and n are the directional cosines.

\begin{figure}
    \centering
    \includegraphics[width=0.99\linewidth,angle=0]{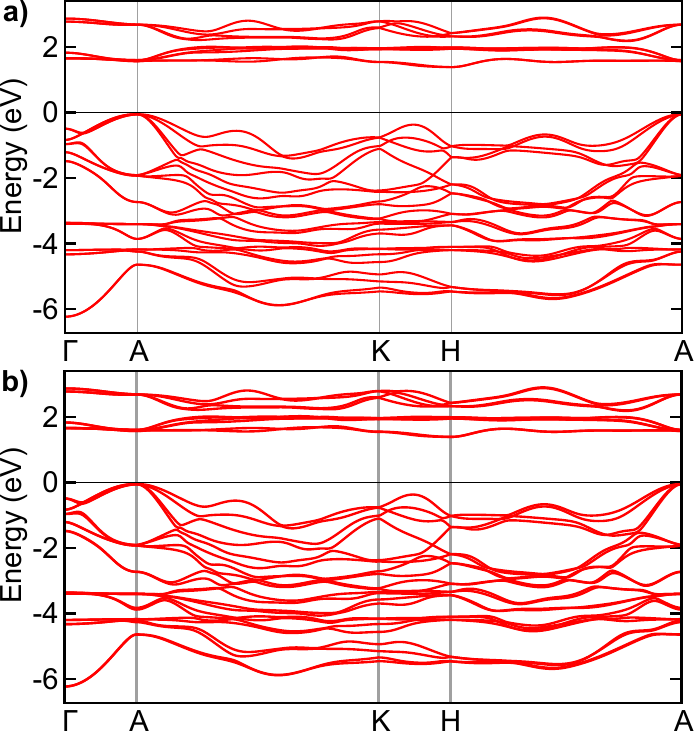}
    \caption{The relativistic altermagnetic band structure calculated from the NM Hamiltonian upon including the requisite on-site spin splitting and SOC. (a) Band structure with electric field orthogonal to the N\'eel vector. (b) Band structure with electric field along the N\'eel vector. The zero of the energy scale is set at the top of the valence band.}
    \label{fig:BS_Mag}
\end{figure}

Once we obtain the non-magnetic Hamiltonian for the generic angle with and without the electric field, we add the magnetic moment along the easy axis and the SOC using the code developed by our group\cite{benny2026staggereddzyaloshinskiimoriyacantingangle,jskol_SOC_Code_V1_2025}. This ensures the correct magnetic symmetry of the Hamiltonian, including the relativistic spin-momentum locking, and consequently the correct symmetry for the AHC. 
The non-magnetic Hamiltonian already contains the crystal symmetries of the NiAs space group. Once the local spin-splitting is added, the altermagnetic spin-splitting will appear. We add the on-site terms as the on-site spin-splitting $\Vec{h}(\theta_S,\phi_S)$ dependent on the angles and the SOC $\lambda_{Mn}$=20 meV\cite{Miarro2024} and $\lambda_{Te}$=500 meV\cite{Autieri:2017_PM}.
\begin{equation}
H^{split}+H^{SOC}= \sum_i(-\Vec{h}(\theta_S,\phi_S)_i\cdot\Vec{S}_i+\lambda_i\Vec{L}_i\cdot\Vec{S}_i) 
\end{equation}
where the modulus of the on-site spin-splitting of Mn that we have used is 5 eV, which is needed to reproduce the experimental band gap of around 1.3 eV\cite{ZANMARCHI19672123}.
More in detail, the maximum of the valence band is at A, while the minimum of the valence band is at H among the high-symmetry points with a band gap of 1.301 eV. The maximum of the valence band at A agrees with what is reported by DFT calculations\cite{junior2023sensitivity}. 
We can compare the band structure obtained using this approach with the DFT band structure shown in Figure 3(a) of reference \onlinecite{g32j-hnvz}. Overall, we observe fair agreement, except at the K point, where the bottom of the conduction band lies only 0.7 eV above the top of the valence band. This is inconsistent with the experimentally measured band gap of 1.3 eV\cite{ZANMARCHI19672123}. This discrepancy suggests that the DFT prediction at the K point, where it differs from our results, is likely inaccurate.

After adding the lattice polarization P$_x$ and P$_y$, the relativistic altermagnetic band structures obtained in this way are reported in Fig. \ref{fig:BS_Mag}(a) and \ref{fig:BS_Mag}(b), respectively. By eye, it is not possible to see the difference between the two noncentrosymmetric cases and the centrosymmetric case.

\begin{figure}
    \centering
    \includegraphics[width=0.99\linewidth,angle=0]{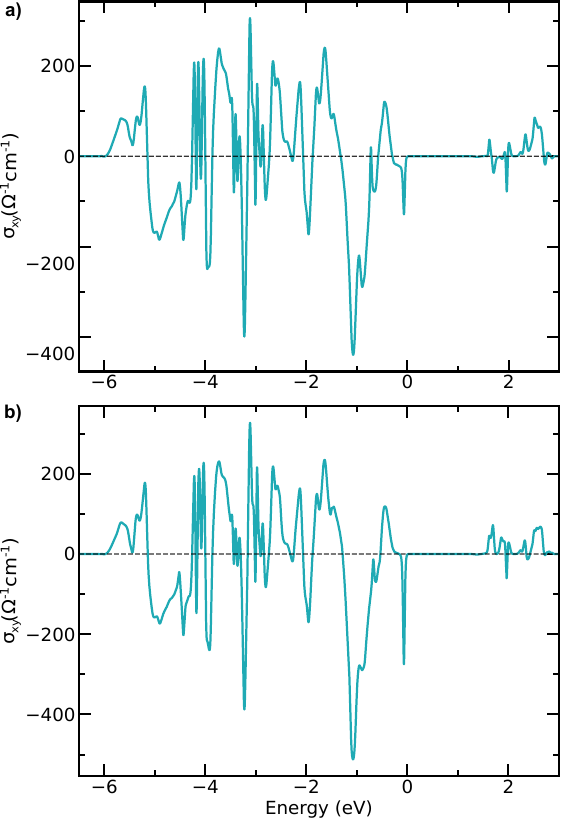}
    \caption{Anomalous Hall conductivity $\sigma_{xy}$ as a function of the energy between -6.5 eV and +3.0 eV. (a) $\sigma_{xy}$ for the electric field orthogonal to the N\'eel vector. (b) $\sigma_{xy}$ for the electric field along the N\'eel vector. The zero of the energy scale is set at the top of the valence band.}
    \label{fig:AHC_Ex_Ey}
\end{figure}

\begin{figure}
    \centering
    \includegraphics[width=0.99\linewidth,angle=0]{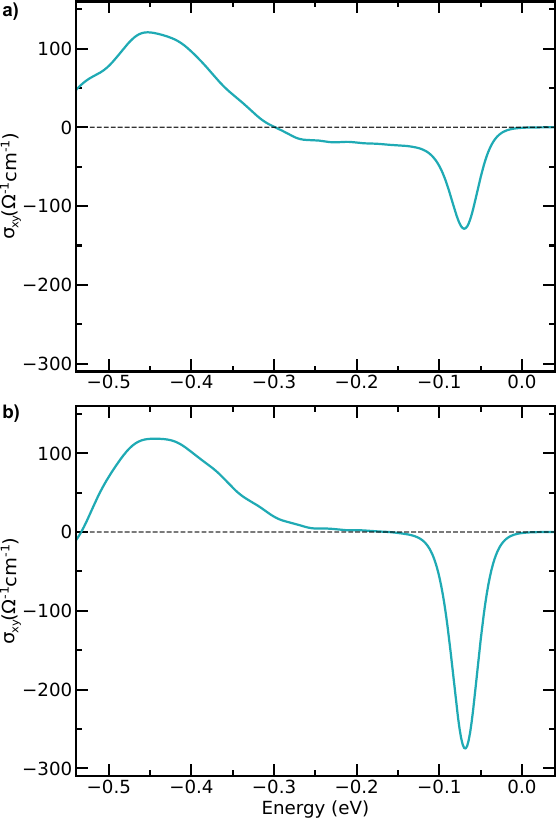}
    \caption{Anomalous Hall conductivity $\sigma_{xy}$ as a function of the energy between -0.5 eV and 0 eV. (a) $\sigma_{xy}$ for the electric field orthogonal to the N\'eel vector. (b) $\sigma_{xy}$ for the electric field along the N\'eel vector. The zero of the energy scale is set at the top of the valence band.}
    \label{fig:AHC_Ex_Ey_zoom}
\end{figure}

\subsection{AHC for different lattice polarizations}
\label{subsec:AHC}

Here, in this section, our goal is to address the effect of the lattice polarization on the AHC as a function of the direction of the electric field in the prototypical altermagnet MnTe.
For the intrinsic AHC, the Hall conductivity tensor obeys the Onsager relation $\sigma_{xy}$=-$\sigma_{yx}$, which has been shown to hold in MnTe\cite{PhysRevLett.130.036702}. This relation follows from the antisymmetric nature of the Hall conductivity tensor. 
Experimentally, the conductivity $\sigma_{xy}$ and the resistivity R$_{xy}$ can be measured as a function of the orientation of the applied electric field\cite{bangar2025interplayaltermagneticordercrystal}, and if this antisymmetric property holds, one should observe a periodicity of 90$^\circ$ with respect to the electric field.
This property may be violated in circumstances where $\sigma_{xy}$ varies as a function of electric field orientation. One possible explanation for this is the consideration of having different extrinsic contributions for different angles of the electric field; another explanation is the effect of the electric field, which, in a semiconductor like MnTe, could displace the anions and cations in inequivalent ways as happens here for the $x$- and $y$-direction. Indeed, of particular interest is the case of a doped ionic semiconductor, such as MnTe. These systems have lattice polarization (atoms) and electrostriction (volume) for which the electric field slightly shifts the ions within the unit cell. This induces lattice polarization through tiny atomic displacements smaller than 1 \AA, which are reversible and exhibit no long-range diffusion.
In the standard metallic cases, $\sigma_{xy}$ is almost constant given that external factors are not involved. Hence, the Onsager relation is valid. However, if other factors appear, such as the lattice polarization, then $\sigma_{xy}$ is no longer a constant and can depend on the orientation of the applied electric field, as we will demonstrate here.

\begin{figure}
    \centering
    \includegraphics[width=0.99\linewidth,angle=0]{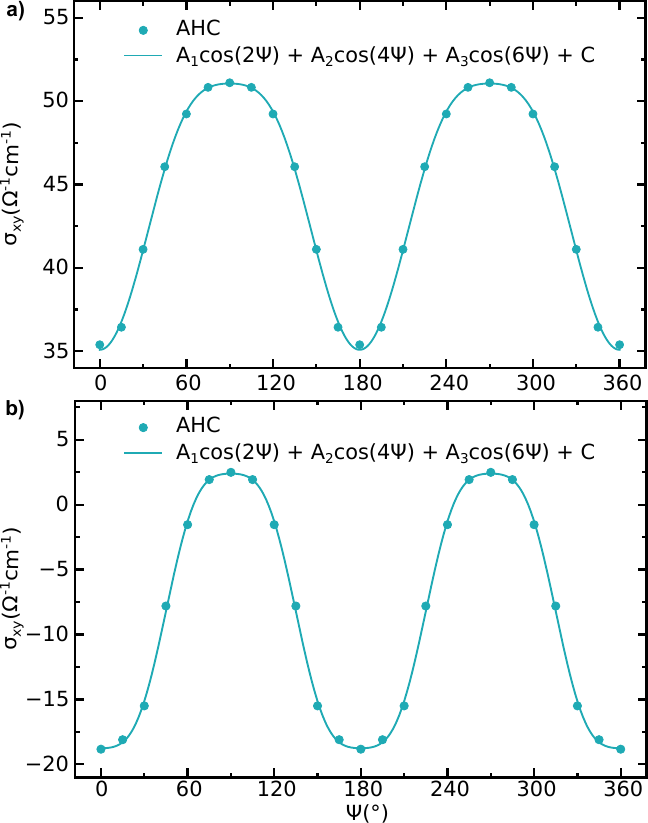}
    \caption{Anomalous Hall conductivity $\sigma_{xy}$ as a function of the angle between the electric field and the N\'eel vector, $\psi$. The Fermi level is set at (a) 345 meV and (b) 216 meV below the top of the valence band. The dots represent the calculated points, while the solid line represents the fitting of the AHC with a periodic function in $\psi$.}
    \label{fig:AHC_Psi}
\end{figure}

The Hamiltonian for the generic direction of the electric field is obtained using the following equation:
\begin{equation}
H(\psi)=|\sin(\psi)|^2H(E_x)+|\cos(\psi)|^2H(E_y) 
\end{equation}
where $\psi$ is defined as the angle between the applied electric field and the $x$-axis, consistent with the experiments\cite{bangar2025interplayaltermagneticordercrystal}.
For H(E$_x$) and H(E$_y$), we report the AHC $\sigma_{xy}$ in Fig. \ref{fig:AHC_Ex_Ey}(a) and \ref{fig:AHC_Ex_Ey}(b), respectively. 
The average value of the AHC is much lower than the centrosymmetric case; we attribute this large variation to the sensitive contribution of the nodal line to the AHC.\cite{benny2026magneticallytunablesymmetryenforcednodal}
While the band structures are similar between the two cases, we can note that the conductivities have a similar energy range and some similar peaks, but they differ in some details, offering energy regions where they have a different sign of $\sigma_{xy}$. At the top of the valence band, there is a peak in $\sigma_{xy}$, which is associated with point A in agreement with DFT results\cite{junior2023sensitivity}.
The magnification of AHC between -0.5 and the Fermi level is presented in Fig. \ref{fig:AHC_Ex_Ey_zoom}(a,b) for the electric field along the $x$- and $y$-axis, respectively. We can clearly observe a sign change of the AHC in the region between -200 and -300 meV. This sign change leads to a violation of the antisymmetric property of the Hall conductivity tensor and the Onsager relation.

We analyzed the anisotropic conductivity as a function of the electric field at two values of the energies, which are 345 meV and 216 meV below the top of the valence band. In the first case, the AHCs corresponding to different electric fields differ in magnitude while retaining the same sign. In the second case, the AHCs have opposite signs.
In the first case, we consider the Fermi level to be 345 meV below the valence band maxima. The AHC $\sigma_{xy}$(E$_F$) plotted as a function of $\psi$ is given in Fig. \ref{fig:AHC_Psi}a.
We can fit our results with the function 
$A_1\cos(2\psi) + A_2\cos(4\psi) + A_3\cos(6\psi) + C$. The fitting parameters, $A_1$, $A_2$ and $A_3$, obtained turn out to be -8.101 $\Omega^{-1}\text{cm}^{-1}$, -1.427 $\Omega^{-1}\text{cm}^{-1}$ and 0.113 $\Omega^{-1}\text{cm}^{-1}$ respectively. This reveals that the $2\psi$ component dominates the AHC, while both the $4\psi$ and $6\psi$ components are present but play a minor role in the overall behaviour of the system. Our theoretical results qualitatively agree with the experimental results\cite{bangar2025interplayaltermagneticordercrystal}.
In the second case, when the Fermi level is fixed 216 meV below the top of the valence band, the AHC changes sign upon varying $\psi$ as shown in Fig. \ref{fig:AHC_Psi}b. It can be fitted with the aforementioned fitting function which yields the values -11.67 $\Omega^{-1}\text{cm}^{-1}$, 0.016 $\Omega^{-1}\text{cm}^{-1}$ and 1.092 $\Omega^{-1}\text{cm}^{-1}$ corresponding to $A_1$, $A_2$ and $A_3$, respectively. 

\subsection{Relativistic spin-momentum locking}

The dominant component of MnTe in the NiAs structure is $Q_{yz}$\cite{Fakhredine26,hirakida2025multipoleanalysisspincurrents}.
When we applied the electric field along the y-axis, we observed the Rashba effect, which influences only the component orthogonal to the electric field\cite{D6MH00357E}.
Therefore, the dominant component does not change with respect to MnTe in the NiAs structure. Since the electric field produces only the Rashba term, which does not generate AHC, the only allowed tensor component of the AHC is $\sigma_{xy}$ for all considered cases.

\begin{table}[h!]
\centering
\begin{tabular}{|c|c|c|c|}
\hline
& \multicolumn{3}{c|}{Spin components} \\
\hline
N\'eel vector & $S_x$ & $S_y$ & $S_z$ \\
\hline
$N \parallel y$ & $Q_{xz}$ & $Q_{yz}$ & $Q_{0}$ + $Q_{xx}$-$Q_{yy}$\\
\hline
$N \parallel y$ with E$_y$ & $Q_{xz}$ + $Q_{z}$ & $Q_{yz}$ & $Q_{0}$ + $Q_{xx}$-$Q_{yy}$ + $Q_{x}$\\
\hline
$N \parallel y$ with E$_x$ & $Q_{xz}$ & $Q_{yz}$ + $Q_{z}$ & $Q_{0}$ + $Q_{xx}$-$Q_{yy}$  + $Q_{y}$\\
\hline
\end{tabular}
\caption{Lowest wave symmetry allowed of the RSML of MnTe with NiAs hexagonal structure for the N\'eel vector oriented along the $y$ axis and for the different spin components $S_x$, $S_y$, and $S_z$, with and without electric field. The S$_y$ component is the dominant one for all considered cases.}
\label{tab:CrSb_AM}
\end{table}

\begin{figure*}
    \centering
    \includegraphics[width=\linewidth,angle=0]{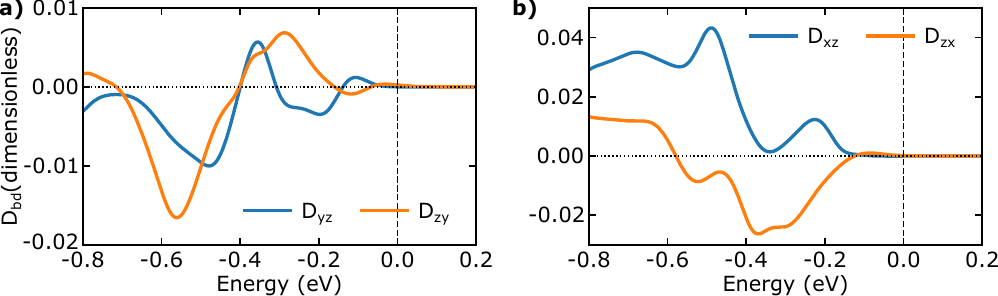}
    \caption{Berry curvature dipole $D_{bd}$, in dimensionless units, as a function of chemical potential for polar distortions along (a) the $x$ direction and (b) the $y$ direction. The energy is measured relative to the Fermi level, marked by the vertical dashed line. For the $P_x$-distorted structure, the symmetry-allowed components are $D_{yz}$ and $D_{zy}$, whereas for the $P_y$-distorted structure, the symmetry-allowed components are $D_{xz}$ and $D_{zx}$.}
    \label{fig:8}
\end{figure*}

\section{Berry Curvature Dipole and Nonlinear Hall effect}
We now discuss the nonlinear Hall response induced by the electric-field-driven polar distortions in altermagnetic MnTe. For both polar-distorted structures considered below, the antiferromagnetic N\'{e}el vector, denoted by \(\mathbf{N}\), is kept along the \(y\)-direction.
The polar distortion breaks inversion symmetry and therefore allows a finite BCD, which provides the intrinsic quantum geometric contribution to the second-order nonlinear Hall response in the relaxation-time approximation. The BCD tensor is defined as\cite{BCD2015Sodemann},

\begin{eqnarray*}
    D_{bd}&=&\sum_n\int [d\mathbf k], v_{n,b}(\mathbf k),\Omega_{n,d}(\mathbf k)\left(-\frac{\partial f_0}{\partial \varepsilon_n}\right),
\end{eqnarray*}

where \(v_{n,b}(\mathbf k)\) is the band velocity, \(\Omega_{n,d}\) is the Berry curvature, $\varepsilon_n$ is the band-dispersion, $f_0$ is the equilibrium Fermi--Dirac distribution  and $n$ being the band index. The first index of \(D_{bd}\) tensor therefore transforms as a polar vector, while the second index transforms as an axial vector. The symmetry classification of the two polar-distorted bulk structures is summarized in Table~\ref{tab:space_groups}. For the \(P_x\)-distorted structure, corresponding to the Mn displacement along the \(x\)-direction, the spatial space group symmetry is of \(Ama2\)-type and the magnetic space group symmetry (MSGs) for \(\mathbf{N}\parallel y\) is \(Ama2.1\), with magnetic point group symmetry (MPGs) \(mm2\). This \(mm2\)-type point symmetry with polar axis along \(x\) allows only the nonvanishing BCD components \(D_{yz}\) and \(D_{zy}\). For the \(P_y\)-distorted structure,, corresponding to the Mn displacement along the \(y\)-direction, the space group symmetry is of \(Amm2\)-type and the MSGs for \(\mathbf{N}\parallel y\) is \(Amm'2'\), with MPGs \(mm'2'\), as listed in Table~~\ref{tab:space_groups}. Since the BCD is even under time reversal, the primed operations impose the same tensor-shape constraints as their unprimed spatial counterparts. Therefore, the \(P_y\)-distorted structure allows only the BCD components \(D_{xz}\) and \(D_{zx}\). The nonlinear Hall current generated by the BCD can be written as \( j_a = \chi_{abc} E_b E_c \), where the second-order conductivity tensor is given by \( \chi_{abc} = \frac{e^3 \tau}{2\hbar^2} \epsilon_{adc} D_{bd} \), with \( \epsilon_{adc} \) the Levi-Civita symbol and \( \tau \) the relaxation time \cite{BCD2015Sodemann}. These nonzero components of BCD therefore determine the allowed nonlinear Hall response.
The dependence of the BCD on the chemical potential is shown in Fig. \ref{fig:8}. The left panel presents the symmetry-allowed components for the \(P_x\)-distorted structure, while the right panel shows the corresponding results for the \(P_y\)-distorted structure, allowing a direct comparison of how both the magnitude and sign of the BCD vary with the direction of the polar distortion.


\section{Conclusion}

In this work, we have investigated the AHC in altermagnetic $\alpha$-MnTe under breaking of inversion symmetry. In the bulk, the N\'eel vector aligned along the $y$-axis breaks the C$_6$ symmetry down to C$_2$. At the (001) surface, the additional symmetry reduction—where the C$_2$ symmetry is further broken, leaving only a mirror plane with respect to $x=0$—leads to new emergent functionalities. As a result, the (001) surface state supports not only a polar dispersion along the $z$-axis, but also additional polar dispersion along the $y$-axis. These polar distortions along the y-axis, which are intrinsically induced by the surface, were recently observed experimentally even in certain regions of bulk MnTe in the form of a local Amm2 orthorhombic phase.\cite{ren2026atomicscaleobservationsymmetrybreaking}

We emphasize that surface-localized polar reconstructions and bulk polar distortions represent distinct physical mechanisms. Here, we investigate both cases independently through slab calculations and periodic bulk calculations, respectively. The bulk results describe the intrinsic Berry curvature driven anomalous Hall response of a uniformly distorted crystal, whereas the slab results account for the influence of surface-localized distortions. Since experiments do not yet uniquely establish whether the relevant polar distortion is bulk-like or confined to the surface, these two calculations should be viewed as complementary limits rather than interchangeable descriptions.

To gain physical insight into these effects of breaking inversion symmetry, we modeled the system by considering the impact of an in-plane electric field in bulk MnTe. Given that MnTe is a doped ionic semiconductor, its electronic structure is also sensitive to lattice polarization under external fields. Our analysis demonstrates that such polarization effects can significantly modify the intrinsic anomalous Hall response. We demonstrate that the lattice polarization can break the crystal symmetry in doped ionic semiconductors, invalidating the antisymmetric properties of the AHC. 

Importantly, the electric-field-induced inversion symmetry breaking introduces a nonlinear AHC to the transport properties that coexists with the anomalous Hall effect. This interplay between intrinsic altermagnetic order and field-driven lattice polarization highlights the complex and tunable nature of transport phenomena in MnTe. We have investigated the nonlinear Hall effect, which can also be present in the recently proposed bulk phases without inversion\cite{wu2025opticalsignaturesnoncentrosymmetricstructural,ren2026atomicscaleobservationsymmetrybreaking}. An interesting direction for future work is the calculation of the quantum metric and, in particular, its distribution in momentum space. Together with the Berry curvature dipole, this would provide a more complete description of the quantum geometry of the electronic bands. The effect of the nonlinear Hall effect due to the electric field along the $z$-axis will be presented somewhere else. 
Overall, our results emphasize that both symmetry breaking and lattice polarization must be considered on an equal footing when describing anomalous transport in altermagnets, particularly in the presence of polar distortions.

\section*{Acknowledgments}

The authors thank C. Ortix,  J. S{\l}awi{\'n}ska, X. Gong, A. Kazakov and C. Chen Ye and for useful discussions.
C. A. was supported by the Polish National Agency for Academic Exchange (NAWA) under the Bekker Programme, grant no. BPN/BEK/2025/1/00244/DEC/1.
This research was supported by the "MagTop" project (FENG.02.01-IP.05-0028/23) carried out within the "International Research Agendas" programme of the Foundation for Polish Science, co-financed by the European Union under the European Funds for Smart Economy 2021-2027 (FENG).
C. B. was supported by Narodowe Centrum Nauki (NCN, National Science Centre, Poland) IMPRESS-U Project No. 2023/05/Y/ST3/00191.
We further acknowledge access to the computing facilities of the Interdisciplinary Center of Modeling at the University of Warsaw, Grant g91-1418, g91-1419, g96-1808, g96-1809 and g103-2540 for the availability of high-performance computing resources and support. We acknowledge the access to the computing facilities of the Poznan Supercomputing and Networking Center, Grants No. pl0267-01, pl0365-01, pl0471-01 and pl0694-01.

\medskip

\appendix


\section{Computational details}

First-principles calculations were performed using the Vienna Ab~initio Simulation Package (VASP)~\cite{kresse1993ab,kresse1996efficiency}, 
within the framework of density functional theory (DFT) and employing the projector augmented-wave (PAW) method~\cite{kresse1999ultrasoft}. 
The exchange--correlation potential was described using the generalized gradient approximation (GGA) with the Perdew--Burke--Ernzerhof (PBE) functional~\cite{perdew1996generalized}. 
A plane-wave cutoff energy of 400~eV was used throughout, and the total energy convergence criterion was set to $10^{-6}$~eV. The values of the Coulomb repulsion and Hund's coupling used are $U = 4.0\,\mathrm{eV}$ and $J_H = 0.97\,\mathrm{eV}$~\cite{PhysRevB.96.214418}, within the rotational invariant approach\cite{liechtenstein1995density}.
To extract the non-magnetic Hamiltonian, we employed the WANNIER90 package, which transforms Bloch states into Wannier functions~\cite{Mostofi:2008_CPC,w90}. The orbital basis used is composed of the d-orbitals of Mn and the p-orbitals of Te, as previously done in the literature\cite{g32j-hnvz}.
We obtained the relativistic altermagnetic Hamiltonian with tunable canting angles using the procedure developed by our group, as described in the literature\cite{benny2026staggereddzyaloshinskiimoriyacantingangle,jskol_SOC_Code_V1_2025}. In this paper, the effect of canting is neglected.
The anomalous Hall conductivity was computed using WannierTools~\cite{wanniertools}. An $181 \times 181 \times 181$ $k$-point mesh was used for these calculations based on the Wannier Hamiltonian. Convergence tests with a denser $201 \times 201 \times 201$ $k$-point mesh yielded only minor changes in the conductivity. 
The Berry curvature dipole calculations were performed using an in-house developed code based on the Wannier tight-binding Hamiltonian obtained from WANNIER90. The BCD tensor was evaluated by integrating over a dense Brillouin-zone mesh of \(801\times801\times501\) \(k\)-points. The thermal smearing in the Fermi--Dirac distribution was considered to be \(k_{\mathrm{B}}T=26\) meV, corresponding approximately to room temperature. Convergence was checked against coarser meshes, and the dense mesh was used for all BCD results presented in the main text.


\bibliography{references}
\end{document}